# STUDY ON ELECTROPOLISHING CONDITIONS FOR 650 MHz NIOBIUM SRF CAVITY

V. Chouhan†, D. Bice, F. Furuta, M. Martinello, M. K. Ng, H. Park, T. Ring, G. Wu, Fermi National Accelerator Laboratory, Batavia, USA
B. Guilfoyle, M. P. Kelly, T. Reid, Argonne National Laboratory, City, USA

## Abstract

The PIP II linear accelerator includes different types of niobium SRF cavities including 650 MHz elliptical low (0.61) and high (0.92) β cavities. The elliptical cavity surface is processed with the electropolishing method. The elliptical cavities especially the low-$\beta$ 650 MHz cavities showed a rough equator surface after the EP was performed with the standard EP conditions. This work was focused to study the effect of different EP parameters, including cathode surface area, temperature and voltage, and optimize them to improve the cavity surface.

## INTRODUCTION

Electropolishing (EP) or buffered chemical polishing (BCP) method is used for processing the inner wall surface of superconducting RF (SRF) cavities used in particle accelerators. The PIP II 800 MeV linear accelerator (linac) includes normal and superconducting accelerating sections to accelerate H$^-$ ions. The superconducting section of the linac uses different types of SRF cavities including half-wave resonator, spoke resonator, and low- and high-β 650 MHz elliptical cavities. The latter two types of cavities are being processed with the EP methods. The EP process is used to make the cavity surface smooth and damage-free. Though extensive studies have been performed on EP of 1.3 GHz cavities and small Nb samples to understand the mechanism of EP [1, 2], limited studies were reported on the large-sized elliptical cavities. This work is aimed to optimize EP parameters for low-β (0.61) 650 MHz (LB650) cavity.

## EP SETUP

EP of 650 MHz cavities were performed with a horizonal EP tool available at Argonne National Lab (ANL). Figure 1 shows a photo of the EP tool with a horizontally assembled 650 MHz 5-cell cavity for EP. EP was performed with a standard cathode and a modified cathode. The standard cathode (cathode A) was designed initially with an extended aluminum pieces, called donut, attached to the cathode pipe. In the modified cathode (cathode-B), the donut length was enlarged to make it almost twice of that in the cathode-A. The schematic of the cathode is shown in Fig. 2. A power supply of 20 V x 750 A was used for the standard EP process. A temporary power supply with the rated power of 30 V x 210 A was used for studying the effect of voltage higher than 18 V, which is the standard voltage applied for EP of Nb SRF cavities. During the EP, the cavity temperature was controlled by spraying water on the exterior of the cavity wall. The cooling system was improved for better control of the cavity temperature.

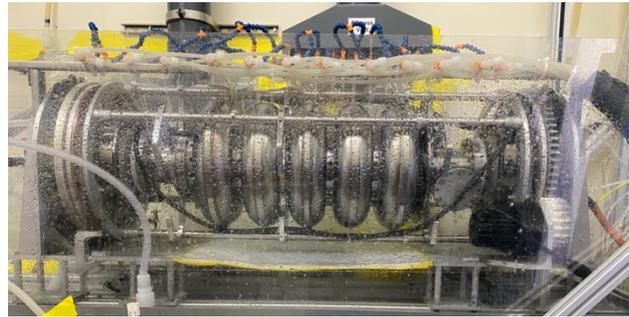

Figure 1: EP tool with an assembled 650 MHz 5-cell cavity.

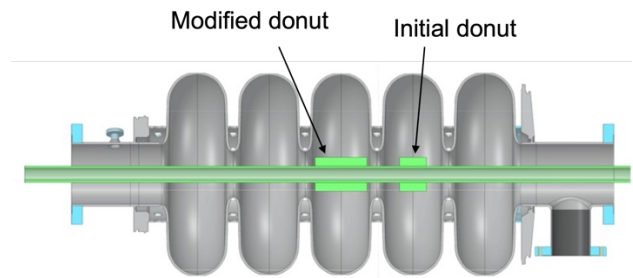

Figure 2: Schematic of the initial (cathode-A) and modified (cathode-B) cathode donuts, which are extended parts on the cathode pipe.

## RESULTS AND DISCUSSION

### Polarization Curves

A polarization curve (*I-V* curve) explains the chemical process occurring on the EP surface as a function of applied voltage [2]. In order to understand the effect of cathode size in the EP process of the LB650 5-cell cavity, *I-V* curves were measured with both types of cathodes. *I-V* curves were obtained while maintaining the cavity temperature to be ~16 ºC. The comparison of *I-V* curves with both cathodes is shown in Fig. 3. The *I-V* curve obtained with cathode-A showed a linear relation between EP current and voltage even up to maximum applied voltage of ~25 V. The plateau region, in which the current remains constant regardless of a change in applied voltage, was not seen. In contrast to this, the cathode-B resulted in the linear region followed by a current plateau region from 17 V onwards. The voltage, at which the plateau region started, is termed as onset voltage.

Table 1: EP parameters applied to two LB650 cavities for bulk removal.

| Parameters | Cavity | |
| --- | --- | --- |
| | B61C-EZ-103 | B61C-EZ-104 |
| Voltage (V) | 18 | 23 |
| Acid flow (L/min) | 8 | 8 |
| Cavity temperature (ºC) | 22 | 16–22 |
| Cavity temperature in cold EP (ºC) | 13 | Cold EP Not applied |
| Cavity rotation (rpm) | 1 | 1 |

*I-V* curves with cathode-B were also measured at higher cavity temperatures up to 22 ºC. An *I-V* curve measured with cathode-B at a temperature of 22 ºC. The onset voltage for EP at 22 ºC was estimated to be ~20 V, which was higher than that observed at a lower cavity temperature.

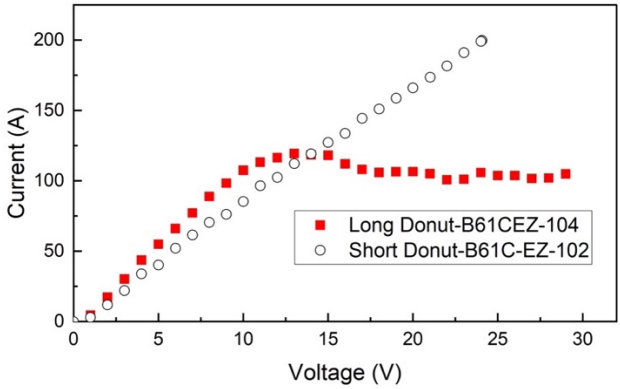

Figure 3: *I-V* curves obtained for 5-cell LB650 cavities with cathode A and B at cavity temperature of ~16 °C.

The *I-V* curves apparently showed that a small cathode surface area could not yield a polishing plateau, which was necessary for EP, even at a significantly higher voltage. With the enlarged cathode, an onset voltage was ~20 V at the cavity temperature of 22 ºC. This onset voltage was higher than the standard onset voltage with a typical value of 12–15 V for 1.3 GHz cavity [1,2]. The larger distance between the cathode and equator surface was supposed to be the cause of the higher onset voltage. However, the distance effect became less pronounce when the cathode surface area was enlarged. A dielectric layer including the oxide layer and diffusion layer on the Nb surface might not be formed in case of the small cathode surface area, large anode-cathode distance, and a higher cavity temperature. A higher cavity temperature might have two effects: (1) It enhances removal rate and reduces thickness of oxide layer on the surface (2) It enhances current and hydrogen gas bubbles that screens the cathode surface to reduce electric field on the Nb surface [1]. An absent or partially formed dielectric layer on EP surface could result in a linear increase of the current with an applied voltage.

## Bulk EP

Bulk EP of two 5-cell LB650 cavities B61C-EZ-103 and B61C-EZ-104 was performed with cathode A and B, respectively. The conditions applied for EP of the cavities are summarized in Table 1. The cavity B61C-EZ-103 was electropolished to remove 150 μm at 22 ºC followed by 10 μm removal in cold EP at a cavity temperature of ~13 ºC. Additional cold EP was performed for removal of 7 μm after 2/0 N-doping was applied to the cavity. A higher voltage of 23 V was applied for EP of B61C-EZ-104 cavity. The voltage for EP was decided based on the result from *I-V* curve measured at 22 ºC so as to conduct EP in the plateau region. Current and temperature profiles during the EP process are shown in Fig. 4. Average currents were recorded to be ~197 and 170 A during EP of B61C-EZ-103 and B61C-EZ-104, respectively. The cavity temperature in EP of B61C-EZ-104 was intentionally kept lower to be ~17 °C. It is because the current oscillation peaks already reached the rated current of the temporary power supply used for EP. The current with the enlarged cathode was seen with oscillations, which can be found in the polishing region only, confirms that EP of the cavity underwent with adequate conditions.

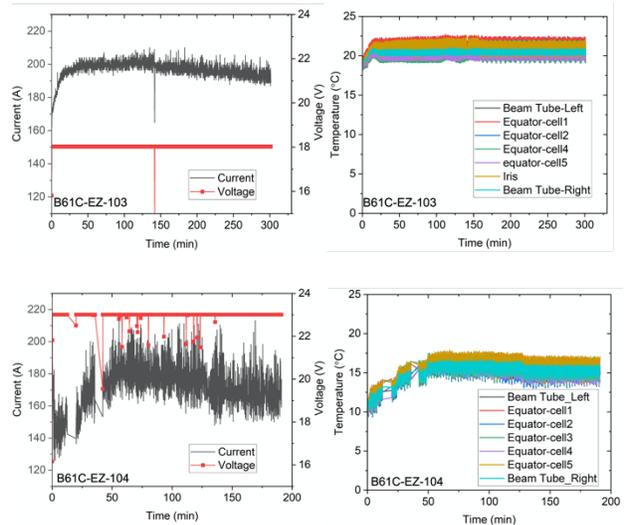

Figure 4: Current and cavity temperature profiles during EP of B61C-EZ-103 (top) and B61C-EZ-104 (bottom).

## Cavity Surface after EP

Equator surface of both cavities was observed with an optical camera. In the case of B61C-EZ-103, a replica of the equator surface in cell-2 was also prepared. The replica was examined with a confocal microscope to obtained detailed surface feature. An optical image of the equator, replica image, and surface profile for the cavity B61C-EZ-103 are shown in Fig. 5. The optical images showed a rough surface of the equator surface. Similar surface feature was observed in all the 5 cells. The replica surface confirms that the electropolished surface was very rough with a grain step height of ~32 µm.

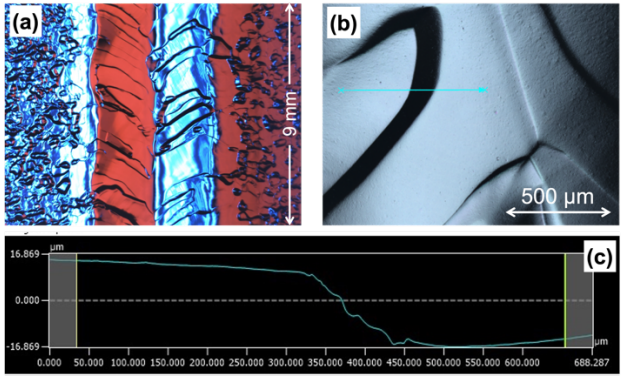

Figure 5: An equator surface of B61C-EZ-103 after the surface processing. (a) Optical image, (b) Confocal microscope image the replica of the equator surface, and (c) surface profile along the line shown in (b).

An optical image of the equator surface from one of cells of B61C-EZ-104 is shown in Fig. 6. The equator surface appeared very smooth like the surface of 1.3 GHz cavity after EP. Iris and beam tube positions were also observed. These positions were also found to be smooth and pit-free. A replica is to be made to know detail on the surface features. The smooth surface confirmed that the modified cathode and EP parameters are in optimum range.

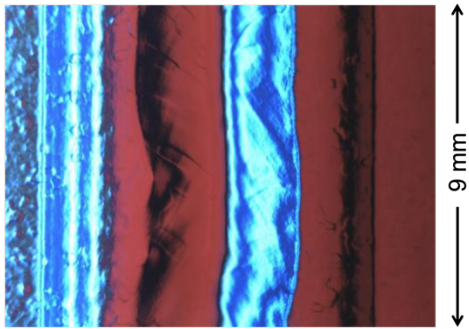

Figure 6: A typical equator surface of the cavity B61C-EZ-104 after bulk EP.

*RF Test*

The cavity B61C-EZ-103 was tested in a vertical cryostat at 2 K. The cavity quenched at ~17 MV/m. This low field was attributed to the sharp grain boundaries with such a large step height on the equator surface. Such a higher step height could enhance a local magnetic field to lower the quench field. The cavity B61C-EZ-104 is to be tested to confirm the effect of the improved equator surface on its SRF performance.

## CONCLUSION

The cavity B61C-EZ-103 was electropolished with the standard cathode and EP parameters. The standard EP resulted in a rough equator surface with a large grain step height of ~32 µm. Such a large step height might be responsible for low-field quench observed in the vertical RF test of the cavity. In order to improve the cavity surface in terms of surface smoothness, the cathode donut size and EP parameters were optimized. The optimized conditions were applied to the cavity B61C-EZ-104 for bulk EP of 120 µm removal. The optimized EP made the entire cavity surface smooth. The optical image of the equator appeared as smooth as the electropolished surface of 1.3 GHz cavity. This study suggests that the standard cathode pipe or a pipe with a short cathode donut, and the standard applied voltage of 18 V are not adequate for large-sized cavities like 650 MHz and 704 MHz cavities.

## ACKNOWLEDGEMENTS

This manuscript has been authored by Fermi Research Alliance, LLC under Contract No. DE-AC02-07CH11359 with the U.S. Department of Energy, Office of Science, Office of High Energy Physics.